\documentclass[twocolumn,showpacs,twoside,superscriptaddress,pra]{revtex4}
\usepackage{amssymb, amsbsy, amsmath, latexsym, dsfont, array, layout, graphicx}
\usepackage{color}

\newcommand{\fig}[1]{Fig.~\ref{#1}}
\newcommand{\tab}[1]{Table~\ref{#1}}

\newcommand{\ket}[1]{\left|{#1}\right\rangle}
\newcommand{\bra}[1]{\left\langle{#1}\right|}

\begin{document}

\title{Quantum walks on circles in phase space via superconducting circuit quantum electrodynamics}
\author{Peng Xue}
\author{Barry C. Sanders}
\affiliation{Institute for Quantum Information Science, University of Calgary, Alberta T2N 1N4, Canada}
\author{Alexandre Blais}
\author{Kevin Lalumi\`ere}
\affiliation{D\'epartement de Physique et Regroupement Qu\'eb\'ecois sur les Mat\'eriaux de Pointe, Universit\'e de
Sherbrooke, Sherbrooke, Qu\'ebec, Canada, J1K 2R1}

\date{\today}

\begin{abstract}
We show how a quantum walk can be implemented for the first time in a
quantum quincunx created via superconducting circuit quantum electrodynamics (QED),
and how interpolation from quantum to random walk is implemented by controllable
decoherence using a two resonator system.
Direct control over the coin qubit is
difficult to achieve in either cavity or circuit QED,
but we show that a Hadamard coin flip can be effected via direct driving of the
cavity, with the result that the walker jumps between circles in phase space but still exhibits quantum walk behavior over
15 steps.
\end{abstract}


\maketitle

\section{Introduction}

The quantum walk (QW) is important in physics as a generalization of the ubiquitous random walk (RW), which underpins
diffusion and Brownian motion. QWs are important in quantum algorithm research~\cite{Aharonov} because they exponentially
speed up hitting time in glued tree graphs~\cite{Chi03}. Although realization of a QW by a quantum quincunx, analogous to the
quincunx (or Galton Board) for realizing the RW, has been proposed in ion traps~\cite{Milburn} and cavity quantum
electrodynamics (QED)~\cite{San03,DHZ04}, the former requires cooling of ions to the center-of-mass motional ground state, and
the latter hitting the atom with pulses that do not drive the cavity at all: these obstacles prevent quantum walks from being
realized under foreseeable experimental conditions (a classical optical simulation of a quantum quincunx has been
performed~\cite{Bou99} but cannot be a proper QW without complementarity~\cite{Ken05}).

Here we devise a quantum quincunx that can realize a QW in the laboratory for the first time by: (i)~employing the
Jaynes-Cummings model~\cite{JC63} to generalize the Hadamard transformation for coin flipping by directly driving the cavity
rather than the atom; (ii)~developing a theory of QWs on many circles in phase space (PS) rather than on a single circle as a
consequence of generalizing the Hadamard transformation; (iii)~optimizing the protocol by having the duration of the
generalized Hadamard transformation depend on the time-dependent mean photon number in the cavity (with detailed theory to
appear elsewhere~\cite{XS08}); (iv)~implementing in a superconducting circuit QED system~\cite{Bla04,Wal05} with a
two-level Cooper Pair Box (CPB) serving as the quantum coin and a coplanar transmission line resonator with a single mode as
the quantum walker; (v)~introducing a double resonator scheme that can control decoherence while simultaneously enabling
strong coupling between the CPB and the microwave field and permitting fast readout; and (vi)~using the Holevo standard
deviation as a measure of phase spreading and showing that the rate of spreading can be tuned by controllable decoherence to
observe the quadratic enhancement of phase spreading for the QW vs RW.

In our scheme the QW is executed with indirect flipping of the coin via directly driving the
cavity and allows controllable decoherence over circles in PS.
Because the walker is directly driven rather than the coin,
photon number is no longer conserved, and
the walker jumps between circles in PS;
however, a signature of quantum walking on circles in PS is evident in both the
time-dependent phase distribution of the walker as well as via direct homodyne measurements to obtain the quadrature
phase (QP) distribution for the walker.
This signature is scaling of the standard deviation~$\sigma$ (which measures the
walker's spreading) that is linear in time~$\propto t$ for the QW, and whose power decreases with increasing decoherence until
attaining the classical RW scaling $\propto\sqrt{t}$ for full decoherence. This controllable decoherence is achieved by
introducing a second low-$Q$ resonator to obtain fast readout~\cite{Yale}.

\section{Background}

To understand the QW in PS, it is helpful to first understand the RW in PS. The walker is a mode of the resonator, hence is
equivalent to a harmonic oscillator, which can be described by its position~$x$ and momentum~$p$. If the oscillator's energy
$E\sim (\omega^2x^2+p^2)/2$ for unit mass and frequency~$\omega$, then the oscillator follows a periodic circular trajectory
of radius~$\sqrt{E}$ in PS with physical oscillatory motion
\begin{equation}
	x(t)=\sqrt{E}\cos\omega t.
\end{equation}
This oscillator can be modified to
execute a RW on a circle in PS by periodically applying an impulse that causes it to rotate either clockwise along the circle
in PS by angle~$\Delta\theta$ or counter-clockwise by the same amount, with the choice of $\pm\Delta\theta$ strictly random.
If~$\Delta\theta=2\pi/d$, $d\in\mathbb{N}$, then the walker always remains on a (perhaps rotating) lattice on the circle with
angular lattice spacing~$\Delta\theta$.

We refer to the coordinate in PS~$(x,p)$ as the walker's `location' in PS, and the
coin flip randomness that determine clockwise vs counter-clockwise angular steps implies that the walker's location is
indeterminate hence described by a distribution~$P(x,p)$.

In an {\em ideal} QW on a circle, the walker+coin state is a density operator~$\rho\in\mathcal{B}(\mathcal{H})$ for
$\mathcal{H}=\mathcal{H}_\text{w}\otimes\mathcal{H}_\text{c}$ (with walker space spanned by $d-1$ discrete phase
states~\cite{Milburn,San03},
\begin{equation}
	\mathcal{H}_\text{w}=\text{span}\{|\theta_m=2m\pi/d\rangle\}
\end{equation}
and coin space
$\mathcal{H}_\text{c}=\text{span}\{|0\rangle,|1\rangle\}$), where $\mathcal{B}(\mathcal{H})$ is the Banach space of bounded
operators on $\mathcal{H}$. The walker's phase distribution on the circle is
\begin{equation}
\label{eq:phasedist}
P_\text{w}(\theta)=\langle\theta|\rho_\text{w}|\theta\rangle/2\pi,\;
    \rho_\text{w}=\text{Tr}_\text{c}\rho,
\end{equation}
with~$d$ equally spaced values of $\theta=\theta_m$.
However, here the walker is following a circular trajectory in PS so the QW's Hilbert space is
$\mathcal{H}_\text{w}=\text{span}\{|n\rangle; n\in\mathbb{N}\}$, with $|n\rangle$ a Fock state
($\hat{n}$ eigenstate).

\begin{figure}[tbp]
   \includegraphics[width=8.5cm]{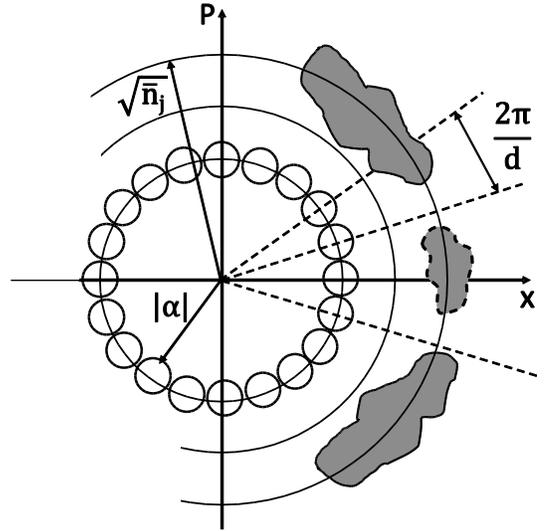}
   \caption{PS diagram depicting circles of fixed radius~$\sqrt{\bar{n}_j}$ for
   circle~$j$. The circle of radius~$|\alpha|$ depicts $1/e$ contours of the
   Wigner function for coherent states separated by angular spacing $2\pi/d$ to
   make them distinct. A generic Wigner function $1/e$ contour (boundary of shaded areas) is depicted  for
   circle~$j$ with mean $\langle\hat{n}\rangle=\bar{n}_j$.
   Solid boundaries represent contours
   around positive peaks, while dashed boundaries are for contours around negative peaks.
   }
   \label{fig:blobby}
\end{figure}

Alternatively the generalized position representation~$\{|x\rangle_\varphi\}$ can be used with
$|x\rangle_\varphi$ an eigenstate of $\hat{x}\cos\varphi+\hat{p}\sin\varphi$, for $\hat{x}$, $\hat{p}$ the canonical
operators satisfying $[\hat{x},\hat{p}]=\mathrm{i}$ ($\hbar\equiv 1$).  The QP distribution is
\begin{equation}
P_\varphi(x)=_\varphi\!\!\langle x|\rho_\text{w}|x \rangle_\varphi
\end{equation}
with $\varphi$ the phase of a local oscillator. A convenient correspondence between the classical and quantum PS trajectory of
the walker is provided by the Wigner quasiprobability distribution
\begin{equation}
\label{eq:Wxp}
    W(x,p)=\int_{-\infty}^\infty \frac{\text{d}y}{2\pi} \text{e}^ {\mathrm{i}py}
        \langle x-y/2|\rho_\text{w}|x+y/2\rangle
\end{equation}
whose marginal distributions are $P_\varphi(x)$. Our scheme for realizing the first experimental QW builds on the cavity QED
quantum quincunx~\cite{San03}, which alternately applies a coin flip Hadamard gate
\begin{equation}
	H=\left|+\right\rangle\langle0|+|-\rangle\langle 1|,
\end{equation}
for
\begin{equation}
	|\pm\rangle=(|0\rangle\pm |1\rangle)/\sqrt{2},
\end{equation}
followed by a
rotation of the walker's state in PS by $\pm\Delta\theta$ with the sign depending on the coin state (with
$\hat{n}=\hat{a}^\dagger\hat{a}$ for $\hat{a}=\left[\omega\hat{x}+i\hat{p}\right]/\sqrt{2\omega}$):
\begin{equation}
F=\exp\left(\mathrm{i}\hat{n} \hat{\sigma}_z\Delta\theta\right).
\end{equation}
The initial state of the walker is a coherent state $|\alpha\rangle$:
$\alpha\in\mathbb{R}$ and
\begin{equation}
	\langle n|\alpha\equiv\sqrt{\bar{n}_0}\rangle=\sqrt{\text{e}^{-\bar{n}_0}\bar{n}_0^n/n!}.
\end{equation}
Fig.~\ref{fig:blobby} depicts PS, including how $d$ is chosen given an initial walker state~$|\alpha\rangle$~\cite{San03}:
\begin{equation}
\bar{n}+\sqrt{\bar{n}}<d<2\pi\sqrt{\bar{n}}.
\end{equation}
With this choice, the walker's angular step size is large enough to ensure
enough distinguishability to yield a circular QW signature.

The simplicity of this model is rooted in the commutativity of $(FH)^N$ with $\hat{n}$, hence a constant of motion; thus the
walker's distance from the PS origin $\sqrt{\bar{n}}$ is fixed. Unfortunately alternating between $F$ and $H$ is not practical
in circuit QED.  For the QW to be realized, and also to have controlled decoherence, a time-dependent driving field for the
resonator is needed. As we see below, the sacrifice is that $[FH,\hat{n}]\neq0$, but for realistic circuit QED conditions, the QW on
circles is (surprisingly) evident provided that the step-by-step $H$ pulse duration is judiciously chosen.

The goal is to observe the spreading of the QW's phase distribution, and the signature of the QW is that this
spread is linear in time: specifically the standard deviation for the phase distribution satisfies $\sigma\propto t$ for the
QW, whereas $\sigma\propto \sqrt{t}$ for the RW, and the QW can be tuned continuously from the RW by controlling decoherence.
As phase is periodic, the usual root-mean-square approach to~$\sigma$ is problematic; instead we employ the Holevo standard
deviation~\cite{Holevo}
\begin{equation}
\label{eq:Holevo}
    \sigma_\text{H}=\sqrt{|\langle \text{e}^{\mathrm{i} \phi}\rangle|^{-2}-1}
\end{equation}
for
\begin{equation}
\langle \text{e}^{\mathrm{i} \phi}\rangle=\int_0 ^{2\pi} d\phi P(\phi) \text{e}^{\mathrm{i}\phi}
\end{equation}
with respect to any phase
distribution $P(\phi)$ (\ref{eq:phasedist}). The Holevo standard deviation
is equivalent to the root-mean-square definition for small spreading on the circle, and $\sigma_\text{H}$
naturally quantifies dispersion over all
$\phi\in[0,2\pi)$\cite{Wis97}. Thus, for sufficiently short times (less than the time that phase distribution spreads
over a significant fraction of the circle in PS), if the relation between phase spreading on the circle with time is
a power law, then
\begin{equation}
\label{eq:logsigmalogt}
    \ln\sigma_\text{H}= \varsigma\ln t +\xi
\end{equation}
with~$\varsigma=1$ for the QW and $\varsigma=1/2$ for the RW.

\section{Superconducting circuit quantum electrodynamics}

The circuit QED Hamiltonian is~\cite{Bla04}
\begin{equation}
\label{eq:H}
    \hat{H}=\hat{H}_\text{JC}+\hat{H}_\text{d},\;
    \hat{H}_\text{d}=\epsilon(t)\left(\hat{a}^\dagger \text{e}^{-\mathrm{i}\omega_\text{d}t}
            +\hat{a}\text{e}^{\mathrm{i}\omega_\text{d}t}\right)
\end{equation}
for $\hat{H}_\text{d}$ the time-dependent driving field Hamiltonian
and
\begin{equation}
\hat{H}_\text{JC}=\omega_\text{r}\hat{n}+\omega_\text{a}
        \hat{\sigma}_z/2+g(\hat{a}^\dagger\hat{\sigma}_-+\hat{a}\hat{\sigma}_+)
\end{equation}
the Jaynes-Cummings (JC) Hamiltonian with $\omega_\text{a}$ and $\omega_\text{r}$ the coin and resonator frequencies,
respectively, and $g$ the coupling strength. It is sufficient to let $\epsilon(t)$ be a square wave so $\epsilon$ is a
constant ($\epsilon=0$ when the field is off). In the dispersive regime,
\begin{equation}
|\Delta|=|\omega_\text{a}-\omega_\text{r}|\gg g,
\end{equation}
and in a frame rotating at $\omega_\text{d}$ for the qubit and the resonator, $\hat H$
can be replaced by the effective Hamiltonian
\begin{align}
\label{eq:Heff}
	 \hat{H}_\text{eff}=&\chi\hat{n}\hat{\sigma}_z/-\delta_\text{da}\hat{\sigma}_z/2
			\nonumber \\ &
		-\delta_\text{dr}\hat{n}+\Omega_R\hat{\sigma}_x/2+\epsilon(\hat{a}^\dagger + \hat{a})
\end{align}
with
\begin{equation}
\delta_\text{da}=\omega_\text{d}-\omega_\text{a}, \delta_\text{dr}=\omega_\text{d}-\omega_\text{r},
\end{equation}
\begin{equation}
\label{eq:Rabi}
	\Omega_R=2g\epsilon/\delta_\text{dr}
\end{equation}
the Rabi frequency, and
\begin{equation}
\chi=g^2/\Delta
\label{eq:cavity pull}
\end{equation}
the cavity pull of the resonator.

The first term in the above expression effects the coin-induced walker phase shift.
The atom transition is an ac-Stark shifted by $g^2\hat{n}\Delta$.
To implement
\begin{equation}
\hat{H}=\exp\left[\mathrm{i}t_\text{H}\Omega_\text{R}\hat{\sigma}_x/2\right]
\end{equation}
on the coin, we choose
\begin{equation}
\omega_\text{d}=2\bar{n}g^2/\Delta-2g\epsilon/\Delta+\omega_\text{a}
\end{equation}
with pulse duration $t_\text{H}=\pi/2\Omega_R$ and $\Omega_\text{R}$ a function of average photon number
\begin{equation}
	\bar{n}(t)=\text{Tr}(\hat{n}\rho_\text{w}).
\end{equation}
The free evolution
\begin{equation}
	\exp{(-\mathrm{i}\hat{H}_\text{eff}t)}
\end{equation}
continues even when the driving field is off $(\epsilon=0)$. For time~$\tau$
between $H$-pulses, the walker steps through an angle
\begin{equation}\Delta\theta\approx\pm g^2(\tau+t_\text{H})/\Delta.\end{equation}

Whereas the ideal QW conserves photon number,
Eq.~(\ref{eq:Heff}) violates this,
which we interpret as the walker wandering between circles in PS.
Circles have radii $\sqrt{\bar{n}(t)}$, and
$\omega_\text{d}$ and $t_\text{H}$ adjusted due to $\bar{n}(t)$ to ensure that the angular step size
$\Delta\theta$ is
constant regardless of how far the walker is from the PS origin. The mean photon number $\bar{n}(t)$ can be calculated as follows~\cite{XS08}.
We begin by solving the Schr\"odinger equation
\begin{equation}
\frac{\text{d}\ket{\varphi}}{\text{d}t}=-\mathrm{i}\hat{H}_\text{eff}\ket{\varphi}
\end{equation}
from time $t=0$ to $t=N(t_\text{H}^{(0)}+\tau)$,
for $N$ the number of steps
\begin{equation}
t_\text{H}^{(0)}=\pi\left[\Delta+2(|\alpha|^2+1)\chi-2g\epsilon/\Delta\right]/4g\epsilon
\end{equation}
and beginning with the initial state
\begin{equation}
\ket{\varphi_\text{o}}=(\ket{0}+\mathrm{i}\ket{1})\ket{\alpha}/\sqrt{2}.
\end{equation}

Figure~2 plots
$\bar{n}(t)$ for $\alpha=3$, $d=21$ and realistic system parameters
\begin{equation}
	(\omega_\text{a},\omega_\text{r}, g,\epsilon)/2\pi=(7000, 5000, 100, 1000)\text{MHz}.
\end{equation}
It is evident in the figure that the mean number oscillates
and then settles down during the free evolution so the walker is concentrated
on a circle of squared radius
\begin{equation}
\bar{n}_j=\left(t_\text{H}^{(0)}+\tau\right)^{-1}\int_{(j-1)(t_\text{H}^{(0)}
    +\tau)}^{j(t_\text{H}^{(0)}+\tau)}\bar{n}(t)dt
\end{equation}
at step~$j$.
The corresponding Hadamard pulse duration~$t_\text{H}^{(j)}$ for each step~$j$ is
\begin{equation}
	 t_\text{H}^{(j)}=\pi\left[\Delta+2(\bar{n}_j+1)\chi-2g\epsilon/\Delta\right]/4g\epsilon
\end{equation}
for $j\in\mathbb{N}$, $\bar{n}_0=9$ and $t_\text{H}^{(0)}=0.01567$ $\mu$s.

Alternatively, to compensate for photon number
fluctuations, it is possible to vary the frequency of the Hadamard pulse rather than its duration.
For large step number~$N$, the photon number
distribution is found numerically to approximate
\begin{equation}
P(n)=|\bra{n}\rho_\text{w}\ket{n}|^2\sim \text{e}^{-\bar{n}}\bar{n}^{n}/n!,
\end{equation}
and the width of the photon number distribution $P(n)$ is closely approximated by
\begin{equation}
\label{eq:deltan}
	\delta n=\sqrt{\langle \hat{n}^2 \rangle-\langle \hat{n} \rangle^2}\approx \sqrt{\bar{n}}
\end{equation}
in the numerical simulations.
Thus, number spreading is negligible, and the walker can be regarded as indeed
being concentrated in the locality of one circle in PS~\cite{XS08}.
As the walker's initial state is a coherent state with a Poissonian number distribution,
the fact that the width remains Poissonian, and the amplitude is relatively constant, indicates that the initially well localized walker continues to be localized with respect to amplitude in the phase space over time.

\section{Localization of walker in phase space}

We observe from numerical simulations that the walker's location in phase space is
effectively localized to a circle in phase space with radial width given by~(\ref{eq:deltan}).
Here we explain why this confinement to the vicinity of a circle in phase space with radius
$\sqrt{\bar{n}}$ is reasonable. This discussion is based on a theoretical analysis of quantum walks
on circles in phase space~\cite{XS08}.

The evolution of the joint walker+coin system is governed by the effective Hamiltonian
(\ref{eq:Heff}). There are five terms on the right-hand side of this Hamiltonian. Let us
understand each term and its effect on the dynamics to see why the walker is localized
to a circle in phase space.
\begin{enumerate}
	\item The first term involves $\hat{n}\hat{\sigma}_z$, which is responsible for
	entangling the evolution of the coin and the walker, effectively to make the walker
	evolve clockwise or counterclockwise at constant amplitude with the orientation
	entangled with the state of the coin.
	\item The second and third terms involve the operators $\hat{\sigma}_z$
	and $\hat{n}$, respectively, which correspond to energies, hence frequencies,
	for the coin and walker.
	\item The fourth term, involving $\hat{\sigma}_x$, is responsible for the Hadamard
	coin flip and is proportional to the Rabi frequency, which is itself
	proportional to the pulsed driving field~$\epsilon$~(\ref{eq:Rabi}).
	\item The fifth term is a displacement involving the operator~$\hat{a}+\hat{a}^\dagger$,
	which pushes the walker off the circle, and is proportional to $\epsilon$.
\end{enumerate}
In making the quantum walk work, the goal is to make the Rabi frequency~$\Omega_\text{R}$
large but keep $\epsilon$ small in order to flip the coin but minimally shift the walker from
the circle.

We have been able to simultaneously achieve the two conditions of large Rabi frequency
and small displacement.
By meeting these two conditions, the evolution is closely approximated by the unitary
evolution
\begin{equation}
	H\otimes D(\mathrm{i}\lambda/\sqrt{2})
\end{equation}
for the quantum walk on circles in phase space~\cite{XS08},
with $\lambda=\sqrt{2}\epsilon t_\text{H}$ the size of the displacements from the circle.
The resultant photon number spread after~$N$ steps is~\cite{XS08}
\begin{widetext}
\begin{equation}
	\delta n\approx \sqrt{\frac{\sqrt{\bar{n}_0}(1+\cos\Delta \theta)}{2}}+\frac{\lambda
	\left[2+\cos\Delta\theta-\cos N\Delta\theta+\frac{\cos(1-N)\Delta\theta}{\sin^2\Delta\theta/2}\right]}{2\sqrt{2}\sqrt{1+\cos\Delta\theta}}.
\end{equation}
\end{widetext}

For large $\bar{n}_0$ and small $\lambda$,
$\delta n\sim \sqrt{\bar{n}}$. Hence, in the asymptotic large mean photon number $\bar{n}$ limit, the reduced walker state has support almost entirely from coherent states with amplitude $\sqrt{\bar{n}}$. This means that the joint state of the walker+coin can be regarded approximately as an entanglement of a walker in superpositions of coherent states with the coin state. This approximation guarantees that the walker can be regarded as being localized to one circle in phase space with a Poissonian spread in photon number that does not increase significantly over time provided that the phase steps are small and $\bar{n}$ is large.



\begin{figure}[tbp]
   \includegraphics[width=8.5cm]{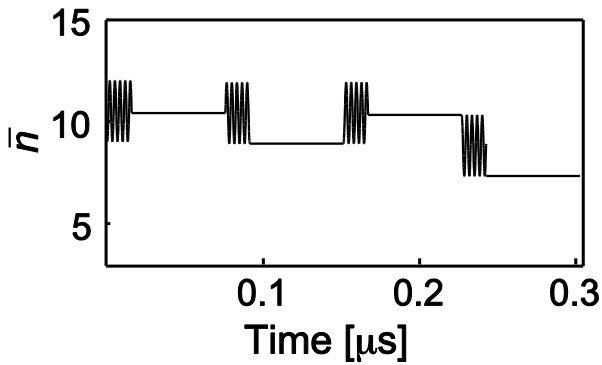}
   \caption{Average photon number~$\bar{n}$ vs evolution time~$t$ ($\mu$s) with period~$(t_\text{H}^{(0)}+\tau)$.
   }
   \label{fig:meanphoton}
\end{figure}

\section{Open system and measurement}

Coupling to additional uncontrollable degrees of freedom leads to energy relaxation and dephasing in the system. In the
Born-Markov approximation, these effects can be characterized by a resonator photon leakage rate~$\kappa$ (determined at fabrication time by the resonator input and output coupling capacitances), an energy relaxation rate $\gamma_{1}$, and a
pure dephasing rate $\gamma_{\phi}$ for the qubit. The open system thus evolves according to
\begin{equation}
\label{eq:mastereq}
    \dot{\rho}=-i\left[\hat{H}_\text{eff},\rho\right]+\kappa
        \mathcal{D}[\hat{a}]\rho+\gamma_{1}\mathcal{D}[\hat{\sigma}_-]\rho
        +(\gamma_{\varphi}/2)\mathcal{D}[\hat{\sigma}_z]\rho,
\end{equation}
with
\begin{equation}
	\mathcal{D}[\hat{L}]\rho\equiv (2\hat{L}\rho \hat{L}^\dagger-\hat{L}^\dagger \hat{L}\rho
		-\rho \hat{L}^\dagger\hat{L})/2.
\end{equation}
Relaxation and dephasing of a charge qubit in circuit QED were reported in~\cite{Wal05} as $T_1=7.3$ $\mu$s and
$T_2=500$ ns. These translate to
\begin{equation}
	\gamma_1/2\pi=0.02 \text{MHz}
\end{equation}
and
\begin{equation}
	\gamma_\phi/2\pi=(\gamma_2-\gamma_1/2)/2\pi=0.31 \text{MHz}.
\end{equation}

The master equation~(\ref{eq:mastereq}) is used to compute~$\rho(t)$ from which
the reduced state of the walker~$\rho_\text{w}$ is obtained and thence the phase
distribution $P_\text{w}(\theta)$~(\ref{eq:phasedist}).
Empirically the phase distribution can be obtained by performing full optical homodyne tomography on
the transmission line resonator to obtain~$W(x,p)$~\cite{Vog89}
from which $P_\text{w}(\theta)$ can be computed and~$\sigma_\text{H}(t)$ thereby determined.
The scaling of~$\sigma_\text{H}$ with~$t$ in Eq.~(\ref{eq:logsigmalogt})
is a convenient empirical signature of the RW vs the QW.

\begin{figure}[t]
   \includegraphics[width=1.0\linewidth]{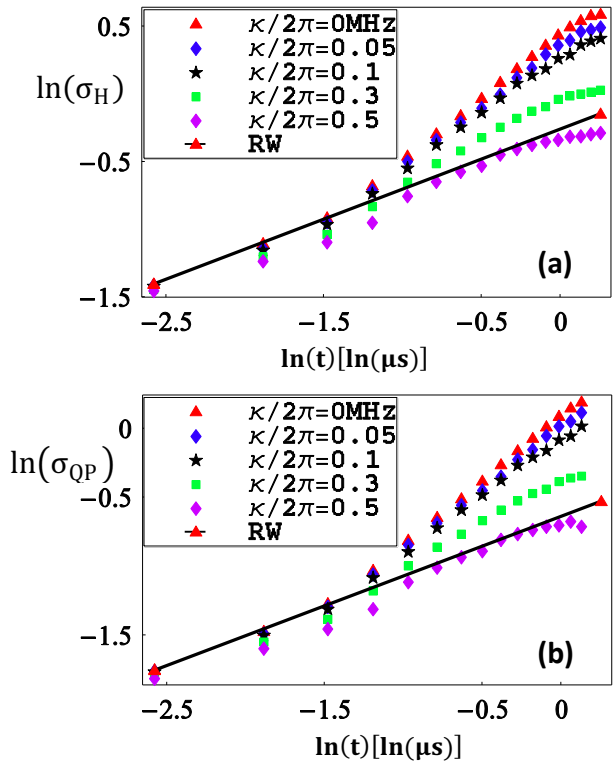}
   \caption{(Color online) Ln-ln plot of walker spread $\sigma$ vs~$t$ ($\mu$s) for
    (a) phase distribution on the circle in PS and
    (b) QP distribution,
    for different $\kappa$ and fixed
    $(g, \gamma_1, \gamma_\phi)/2\pi=(100, 0.02, 0.31)$ MHz.}
    \label{fig:numerical}
\end{figure}

As full tomography is expensive, it would be valuable to observe a QW directly from optical homodyne detection with a single
choice of local oscillator phase~$\varphi$ instead of having to scan over many~$\varphi$ for full tomography. Our simulations
show both~$\sigma_\text{H}(t)$ for $P_\text{w}(\theta)$ and $\sigma_\text{QP}$ for the QP distribution at~$\varphi=0$, with
the choice of~$\varphi=0$ corresponding to having a local oscillator that is in phase with the walker at $t=0$. For our
choices of realistic experimental parameters, the walker's phase distribution spreads from $-\pi$ to $\pi$ on a scale of 15
steps so simulations are limited to fewer than 15 steps before the spreading effectively saturates and our comparison of the
QW vs the RW breaks down.

Simulated evolution of~$\sigma_\text{H}(t)$ and~$\sigma_\text{QP}(t)$ are presented in \fig{fig:numerical}(a,b), respectively.
Corresponding linear regression data presented in \tab{table1} clearly reveals slopes compatible with the characteristic
quadratic \emph{decrease} in phase spreading for \emph{increasing} decoherence of the QW until the transition to the
RW~\cite{San03}.

\begin{table}[t]
 \begin{tabular}{|c|c|c|c|c|c|} \hline
$ \kappa/2\pi \text{(MHz)}$&$ s $&$ \Delta s $&$ \ln\sigma_\text{H}^0 $&$ \Delta\ln\sigma_\text{H}^0 $&$ r $\\
\hline $ 0 $&$ 0.924 $&$ 0.009 $&$ 0.442 $&$ 0.004 $&$ 0.990 $\\ \hline
$ 0.05 $&$ 0.879$&$ 0.013 $&$ 0.362 $&$ 0.006 $&$ 0.991 $\\
\hline $
0.1 $&$ 0.822 $&$ 0.014 $&$ 0.279 $&$ 0.008 $&$ 0.992 $ \\ \hline $ 0.3 $&$ 0.615 $&$ 0.022 $&$ -0.025 $&$ 0.012 $&$ 0.993 $\\
\hline $ 0.5 $&$ 0.447 $&$ 0.030 $&$ -0.309 $&$ 0.017 $&$ 0.990 $\\ \hline
 \end{tabular}
 \caption{The linear regression data
    $\ln\sigma_\text{H}=(s \pm \Delta s)\ln t
    + (\ln\sigma_\text{H}^0 \pm \Delta\ln\sigma_\text{H}^0)$ of the Holevo standard deviation of phase distribution in ln-ln scale for lossy cavities.}
 \label{table1}
\end{table}

\begin{table}[t]
 \begin{tabular}{|c|c|c|c|c|c|} \hline
$ \kappa/2\pi \text{(MHz)}$&$ s $&$ \Delta s $&$ \ln\sigma_\text{QP}^0 $&$ \Delta\ln\sigma_\text{QP}^0 $&$ r $\\
\hline $ 0 $&$ 0.937 $&$ 0.006 $&$ 0.093 $&$ 0.004 $&$ 0.988 $\\ \hline
$ 0.05 $&$ 0.892$&$ 0.009 $&$ 0.016 $&$ 0.006 $&$ 0.989 $\\
\hline $ 0.1 $&$ 0.832 $&$ 0.013 $&$ -0.068 $&$ 0.008 $&$ 0.990 $ \\ \hline $ 0.3 $&$ 0.634 $&$ 0.020 $&$ -0.372 $&$ 0.012 $&$
0.993 $\\ \hline $ 0.5 $&$ 0.453 $&$ 0.034 $&$ -0.677 $&$ 0.019 $&$ 0.990 $\\ \hline
 \end{tabular}
 \caption{The linear regression data $\ln\sigma_\text{QP}=(s\pm\Delta s)\ln t+(\ln\sigma_\text{QP}^0\pm\Delta\ln\sigma_\text{QP}^0)$
 of the standard deviation of QP distribution in ln-ln scale for lossy cavities.}
 \label{table2}
\end{table}

These results show the significance of~$\kappa$ in decoherence from the QW to the RW. Moreover $\kappa$ is much more important
than $\gamma_1$ and $\gamma_\phi$ with respect to the scaling of $\sigma_\text{H,QP}$ with~$t$. The pure dephasing rate
$\gamma_{\phi}$ mainly leads to smearing of the phase distribution and the phase distribution loses its symmetry. Furthermore,
the effect of the energy relaxation rate $\gamma_1$ is small here compared with $\kappa$ because of our realistic choice of parameters.

Unfortunately~$\kappa$ must be low to obtain a QW yet high to allow fast readout. This dilemma is resolved by instead using
two modes~\cite{Wis93}: one resonator (labeled~a) of high-Q and acting as the walker and a second resonator (labeled~b) of
low-Q and used for fast readout~\cite{Yale}. The microwave radiation from resonator~a is coupled into resonator~b, and
measurements ensue on resonator~b. Measurements must be quick on the time scale of the walker's steps, so it cannot be longer
than the time scale between pairs of Hadamard pulses. In the two-resonator system, $\kappa_\text{b}$ has therefore a lower
bound of $O(g^2/\Delta)$.

Due to coupling to resonator~b, the transition frequency of the CPB in the resonator~a and the pure
dephasing rate are changing with the cavity field in resonator~b, so the master equation for resonator~a and the CPB is modified to~\cite{Yale}
\begin{equation}
\label{eq:mastereq'}
\dot{\rho}=-\mathrm{i}[\hat{H_\text{s}},\rho]+\kappa_\text{a}
\mathcal{D}[\hat{a}]\rho+\gamma_{1}\mathcal{D}[\hat{\sigma}_{-}]\rho+\frac{\gamma_{\varphi}+\Gamma_\text{m}}{2}\mathcal{D}[\hat{\sigma}_z]\rho,
\end{equation}
with
\begin{equation}
\hat{H}_\text{s} =\omega_\text{rb}\hat{n}+\frac{\Omega'(t)}{2} \hat{\sigma}_z+\left[g\hat{a}^\dagger\hat{\sigma}_{-}
+\epsilon(t)\hat{a}^\dagger \text{e}^{-\mathrm{i}\omega_\text{d}t}+\text{hc}\right]
\end{equation}
where
\begin{equation}\Omega'(t)=\omega_\text{a}+2\chi_\text{b} (|\alpha_\text{b}(t)|^2+1/2),\end{equation}
and
\begin{equation}
\Gamma_\text{m}=8\chi_\text{b}^2|\alpha_\text{b}(t)|^2/\kappa_\text{b}.
\end{equation}
In these expressions, $\chi_\text{b}=g_\text{b}^2/(\omega_\text{a}-\omega_\text{rb})$ is the cavity
pull of the measurement resonator~b and $\alpha_\text{b}(t)$ is the classical part of the measurement cavity field.
Under these conditions,
the master equation for the CPB and resonator~b~(\ref{eq:mastereq'}) is identical to Eq.~(\ref{eq:mastereq}) except for a parameter change.
It is interesting to note that, in this two-resonator case, qubit dephasing can be tuned by changing the number of photons
injected in the read-out resonator (corresponding to measurement-induced dephasing~\cite{Bla04}).  As a result, it should be
possible to observe the cross-over between the QW and the RW as a function of this tunable dephasing.

In summary we have introduced the following protocol to implement the QW on circles in PS using circuit QED. (i)~Solve the
Schr\"{o}dinger equation from $t=0$ to $t=N(t_\text{H}^{(0)}+\tau)$, to obtain the mean photon number $\bar{n}(t)$
from which the sequence $\{\bar{n}_j\}$ and $\{t_\text{H}^{(j)}\}$ are obtained. (ii)~Prepare a high-Q resonator in its vacuum
state (by simply cooling). (iii)~When the vacuum state is prepared, inject a microwave field with a Gaussian pulse shape and
temporal width $T_\text{G}$ in order to prepare the high-Q resonator in a coherent state $\ket{\alpha}$. (iv) After
$t=2T_\text{G}$, implement the Hadamard pulse on the CPB by injecting a square pulse of frequency $\omega_\text{d}$ into
resonator~b over time scale $t_\text{H}^{(j)}$ for step $j$. (v) Terminate the external driving of the resonator to allow free
evolution over time scale $\tau$. (vi)~Repeat steps (iv) and (v)~$N$ times. (vii)~Perform full tomography on resonator~b by
performing homodyne measurement over many values of $\phi$ and use standard inversion technique on the data.

In circuit QED, the amplifier thermal noise is significant, with more thermal photons present
than the mean resonator photon
number~\cite{Wal05}. However the signal QP distribution and full Wigner function can be obtained from the resultant homodyne
detection statistics by convolving the readout with a thermal function (which is a Gaussian mixture of Gaussian states
centered at the PS origin). The spread of the convolution is determined by the mean thermal photon number, which is typically 20.

The result is expected to be noisy quadrature phase readout, but repetition will yield, on average, the desired linearity
of $\ln \sigma$ vs $\ln t$. We can use a filter algorithm~\cite{Cho99} which takes the inversion formula for the measured
Radon transform of the Wigner function with thermal noise and reconstruct the Wigner function of the corresponding noiseless
signal. Hence we can achieve the noiseless phase distribution and QP distribution with the Wigner function of the noiseless
signal.

\section{Conclusions}

\begin{figure}[t]
   \includegraphics[width=1.0\linewidth]{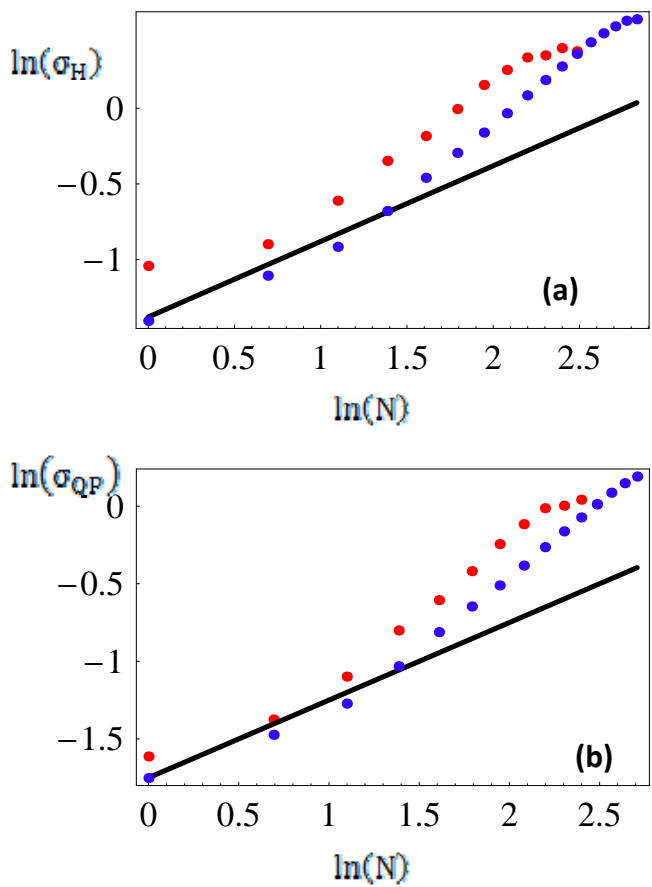}
   \caption{(Color online) Ln-ln plot of walker's phase spread $\sigma$ vs~$N$ (the number of the steps) to $N=15$ with fixed pulse duration $0.0157 \mu s$ for
    (a) phase distribution on the circle in PS and (b) QP distribution for the classical
random walk (solid line) and the QW (red dots) with $\alpha=3$, $d=21$ and typical system parameters
   $(\omega_\text{a},\omega_\text{r}, g,\epsilon)/2\pi=(7000, 5000, 100, 1000)$ MHz. In comparison, the blue dotted lines are for the QW with adaptive pulse durations, which breaks down after 15 steps.}
    \label{fig:without}
\end{figure}

We have shown that the QW on circles in PS can be implemented via circuit QED with and without open systems effects using realistic
parameters and find that the signature of the QW is evident under those conditions. Moreover full tomography may not be
required because direct homodyne measurements over few quadratures reveal an unambiguous QW signature, and the RW can
controllably emerge by tuning decoherence. Our scheme shows how a QW with just one walker can be implemented in a realistic
system for the first time, and controllable decoherence can be performed thereby allowing continuous tunability between the
quantum and classical regimes.

Our scheme allows only a finite number of steps before the quadratic enhancement in
phase spreading breaks down.
As the phase step $\Delta\theta$ must be strictly greater than $1/\delta n$,
the number of steps $N$ has an upper bound because of the desire to avoid wrap-around effects (the walker going around the circle), then $N<2\pi/\Delta\theta$. Therefore, $N <2\pi\delta n\approx 2\pi\sqrt{\bar{n}}$ provides an upper bound on $N$.

Hence the number of steps has an upper bound that is determined by $\bar{n}$. If the mean number of photons in the resonator is increased, so is the allowed number of steps. Physically, however, the mean number of photons cannot be too large
because the dispersive approximation that we exploits
fails for large cavity photon number:
this breakdown occurs for critical photon number~\cite{Bla04,BGB08}
\begin{equation}
\label{eq:ncrit}
	\bar{n}_\text{crit}=\frac{\Delta^2}{4g^2}.
\end{equation}

Eq.~(\ref{eq:ncrit}) yields an upper bound of photon number in the resonator and limits the number of steps the walker can take and exhibit a quadratic enhancement of spreading.
In our example, $\Delta=2000$ MHz and $g/2\pi=100$ MHz so $\bar{n}_\text{crit}=100$.
Here we have treated the case of just nine photons, and seen strong evidence
of a quadratic enhancement of phase diffusion, so the quantum quincunx effectively
works well below this critical photon number where the dispersive approximation breaks down.

In our scheme, the pulse duration~$t_\text{H}$ is adjusted each time according to the predicted mean photon number, but precise control of the Hadamard pulse may be difficult to achieve in experiments.

In Fig. 4 we have shown a simulation of the cases with and without adjusting the duration of the Hadamard pulse sequence. Not adjusting the Hadamard pulse durations still yields a quadratic enhancement of the phase distribution due to the quantum walk, but it breaks down earlier. In our simulation, the breakdown occurs after 10 steps rather than 15 for the adaptive pulse duration. For the first 10 steps, the numerically simulated standard deviation for QP distribution and the Holevo standard deviation in ln-ln scale are respectively shown to be approximately linear in $\ln{N}$: $\ln{\sigma_\text{QP}}=(0.939\pm0.007)\ln{N}+(-2.090\pm0.005)$, the $r$ coefficient is 0.99, and $\ln{\sigma_\text{H}}=(0.890\pm0.006)\ln{N}+(-1.563\pm0.003)$, the $r$ coefficient is 0.96.

\begin{acknowledgments}
This work has been supported by NSERC, MITACS, CIFAR, FQRNT, QuantumWorks and iCORE. We thank Stephen Bartlett, Jay Gambetta,
Steve Girvin, and Rob Schoelkopf for helpful discussions.
\end{acknowledgments}

\end{document}